\documentclass[preprint,amsmath,amssymb]{revtex4-1}
\usepackage{xcolor}
\usepackage{tabu}
\usepackage{titlesec}
\usepackage[utf8]{inputenc}
\usepackage{graphicx,epsfig,epstopdf}
\usepackage{bm}
\usepackage{gensymb}
\usepackage{epstopdf}
\usepackage{amsmath,amsfonts,latexsym}
\usepackage{dcolumn}
\usepackage{bm}
\usepackage{hyperref}
\newcommand{\be}{\begin{equation}}
\newcommand{\en}{\end{equation}}
\usepackage[utf8]{inputenc}
\newcommand{\nn}{\nonumber\\}
\usepackage{breqn}             % These four lines are
\makeatletter                  % used to line-break equations
\let\cat@comma@active\@empty   % automatically
\makeatother                   % use \begin{dmath} ... \end{dmath} for eqns
\begin{document}

\title{Semileptonic decays of charmed mesons to light scalar mesons}%

\author{N. R. Soni}
\email{nrsoni-apphy@msubaroda.ac.in}
\affiliation{Department of Physics, Faculty of Science, \\ The Maharaja Sayajirao University of Baroda, Vadodara 390002, Gujarat, {\it INDIA}.}

\author{A. N. Gadaria}
%\email{angadaria-apphy@msubaroda.ac.in}
\affiliation{Applied Physics Department, Faculty of Technology and Engineering, \\ The Maharaja Sayajirao University of Baroda, Vadodara 390001, Gujarat, {\it INDIA}.}

\author{J. J. Patel}
%\email{jjpatel-apphy@msubaroda.ac.in}
\affiliation{Applied Physics Department, Faculty of Technology and Engineering, \\ The Maharaja Sayajirao University of Baroda, Vadodara 390001, Gujarat, {\it INDIA}.}

\author{J. N. Pandya}
\email{jnpandya-apphy@msubaroda.ac.in}
\affiliation{Applied Physics Department, Faculty of Technology and Engineering, \\ The Maharaja Sayajirao University of Baroda, Vadodara 390001, Gujarat, {\it INDIA}.}

\date{\today}

\begin{abstract}
Within the framework of Covariant Confined Quark Model, we compute the transition form factors of $D$ and $D_s$ mesons decaying to light scalar mesons $f_0(980)$ and $a_0(980)$. The transition form factors are then utilized to compute the semileptonic branching fractions. We study the channels namely, $D_{(s)}^+ \to f_0(980) \ell^+ \nu_\ell$ and $D \to a_0(980) \ell^+ \nu_\ell$ for $\ell = e$ and $\mu$. For computation of semileptonic branching fractions, we consider the $a_0(980)$ meson to be the conventional quark-antiquark structure and the $f_0(980)$ meson as the admixture of $s\bar{s}$ and light quark-antiquark pairs. Our findings are found to support the recent BESIII data.
\end{abstract}

\maketitle

\section{Introduction}
\label{sec:introduction}
Charmed semileptonic decays are important for the study of open flavor hadron spectroscopy in general and heavy quark decay properties in particular. More specifically, the scalar mesons below 1 GeV in final product can provide the key information regarding their internal structure as well as the chiral symmetry in the low energy region of nonperturbative QCD \cite{Jaffe:1976ig}. The internal structure of these mesons is yet to be clearly understood from the theoretical studies attempted so far. %The semileptonic decays of $D$ and $D_s$ mesons to pseudoscalar and vector mesons have been studied by various theory groups and these channels are also investigated experimentally world wide.

Conventionally, the structure of $f_0(980)$ meson was thought to be a bound state of quark-antiquark pair.
CLEO was the first to have studied the semileptonic decays of $D_s \to f_0(980) e^+ \nu_e$ \cite{Ecklund:2009aa,Yelton:2009aa} and recently, BESIII has reported branching fractions for the semileptonic decays of $D^+ \to f_0(980) e^+\nu_e$ \cite{Ablikim:2018qzz} suggesting that the internal structure of bound state of $f_0(980)$ meson may not be the conventional quark-antiquark pair.
These experimental results also indicate that the internal structure of $f_0(980)$ meson could be admixture of lighter quark state and $s\bar{s}$ state. For the $D \to f_0(980)$, the dominant contribution is from the lighter quarks whereas that for the $D_s \to f_0(980)$ channel is from the $s\bar{s}$ counterpart.
Recently, BESIII Collaboration have reported the first ever experimental observation for the semileptonic branching fraction of $D \to a_0(980) e^+ \nu_e$ \cite{Ablikim:2018ffp}.
The notable results on these channels read,
\begin{eqnarray}
\mathcal{B}(D^+ \to f_0(980) e^+ \nu_e, f_0(980) \to \pi^+\pi^-) & < & 2.5 \times 10^{-5} \nonumber \ \ \text{\hfill\cite{Ablikim:2018qzz}}\\
\mathcal{B}(D_s^+ \to f_0(980) e^+ \nu_e, f_0(980) \to \pi^+\pi^-) & = & (0.13 \pm 0.04 \pm 0.01)\%  \nonumber \ \ \text{\hfill\cite{Yelton:2009aa}}\\
\mathcal{B}(D^0 \to a_0(980)^- e^+ \nu_e, a_0(980)^- \to \eta\pi^-) & = & (1.33^{+0.33}_{-0.29} \pm 0.09) \times 10^{-4} \ \ \text{\hfill\cite{Ablikim:2018ffp}}  \\
\mathcal{B}(D^+ \to a_0(980)^0 e^+ \nu_e, a_0(980)^0 \to \eta\pi^0) & = & (1.66^{+0.81}_{-0.66} \pm 0.11) \times 10^{-4} \ \ \text{\hfill\cite{Ablikim:2018ffp}} \nonumber 
%\frac{\Gamma(D^0 \to a_0(980)^-e^+ \nu_e)}{\Gamma(D^+ \to a_0(980)^0e^+ \nu_e)} & = & 2.03 \pm 0.95 \pm 0.06 \ \ \text{\hfill\cite{Ablikim:2018ffp}} \nonumber
\end{eqnarray}
On experimental front, BESIII and other worldwide experimental facilities have reported the most precise results on semileptonic decay of $D_{(s)}$ to pseudoscalar and vector mesons.
From theory point of view, these channels are straightforward to study because the internal structure/quark content of the daughter meson is a typical quark-antiquark system.
But the quark structure of the scalar mesons below 1 GeV has varied explanations (see note on scalar mesons below 2 GeV in Particle Data Group (PDG) Ref. \cite{Tanabashi:2018oca}).
The computation of branching fractions of $D_{(s)}$ mesons decaying to $a_0(980)$ and $f_0(980)$ is highly sensitive to the internal structure of these mesons.
The theoretical approaches so far include lattice quantum chromodynamics \cite{Aubin:2004ej,Na:2010uf,Na:2011mc,Bali:2014pva,Lubicz:2017syv,Lubicz:2018rfs}, QCD sum rules \cite{Bediaga:2003zh,Wu:2006rd,Offen:2013nma,Duplancic:2015zna,Momeni:2019uag}, chiral unitary approach \cite{Sekihara:2015iha} and different quark models \cite{Melikhov:2000yu,Fajfer:2004mv,Fajfer:2005ug,Verma:2011yw,Palmer:2013yia,Cheng:2017pcq}.
There are different ways in which these scalar mesons are studied globally viz. conventional quark-antiquark states \cite{Morgan:1993td,Ke:2009ed,Issadykov:2015iba} and compact multiquark states including diquark-diantiquark \cite{Fariborz:2009cq}, meson-meson composite molecule \cite{Weinstein:1982gc,Weinstein:1990gu,Achasov:1996ei,Branz:2007xp,vanBeveren:2003kd,Pelaez:2003dy,Hooft:2008we,Dai:2012kf,Dai:2017uao}, as well as compact structure of tetra quarks \cite{Achasov:2010fh,Achasov:2012kk,Achasov:2017zhy,Achasov:2018grq,Maiani:2004uc,Maiani:2007iw,Kim:2018zob,Kim:2017yur,Kim:2017yvd}.
In this mass range, one can also consider the possibility of scalar glueball bound state \cite{Narison:1996fm,Minkowski:1998mf}.

In quark-antiquark picture, several theories have been proposed including the constituent quark model \cite{Morgan:1993td}. The semileptonic branching fractions for $D_s \to f_0(980)$ were considered in the light front quark model \cite{Ke:2009ed} and semileptonic as well as rare decays of the $B_{(s)}$ have also been studied in the formalism of covariant quark model \cite{Issadykov:2015iba}.
In composite structure, there are several ways in which the structure of scalar mesons is proposed.
A. H. Fariborz \textit{et al}., have studied the light scalar mesons considering diquark - diantiquark state in the formalism of linear sigma model \cite{Fariborz:2009cq}.
The scalar mesons have been considered to be the bound states of $K\bar{K}$ molecules in the potential model approach \cite{Weinstein:1982gc,Weinstein:1990gu}.
N. N. Achasov et al have considered the scalar meson to be the $K\bar{K}$ molecule in the radiative decays of $\phi$ meson \cite{Achasov:1996ei}.
The assignment of $f_0(980)$ as the molecular structure of $K\bar{K}$ has also been used in the phenomenological Lagrangian approach by studying the strong decays of $f_0(980)$ to $\pi\pi$ and $\gamma\gamma$ channels \cite{Branz:2007xp}.
The multiquark structure was also attempted in the unitarized meson model \cite{vanBeveren:2003kd}, effective field theory \cite{Pelaez:2003dy,Hooft:2008we} as well as chiral perturbation theory \cite{Dai:2012kf,Dai:2017uao}.
L Maiani \textit{et al}., have studied the diquark-diantiquark structure via strong decay of $D_{(s)}$ mesons \cite{Maiani:2004uc,Maiani:2007iw}.
Light scalar mesons are also studied in the framework of tetraquark mixing \cite{Kim:2018zob,Kim:2017yur,Kim:2017yvd}.

Lattice quantum chromodynamics investigations of these scalar mesons are reported employing the four quark  \cite{Wakayama:2014gpa,Alexandrou:2017itd} and diquark-diantiquark pictures \cite{Alford:2000mm}.
The transition form factors for the decays with scalar mesons as daughter products are also computed in the framework of QCD sum rules \cite{Cheng:2005nb,Aliev:2007uu,Wang:2009azc} and light cone sum rules \cite{Colangelo:2010bg,Shi:2017pgh}.
Cheng \textit{et al}., studied the transition form factors and semileptonic branching fractions for the channel $D \to a_0(980)$ in the light cone sum rule approach where they considered the $a_0(980)$ to be the conventional quark antiquark state \cite{Cheng:2017fkw}.
The transition form factors are also determined in the light front quark model considering them as a four quark state \cite{Verma:2011yw}.
R. L. Jaffe has considered the diquark-diantiquark structure of these mesons in the formalism of MIT bag model \cite{Jaffe:1976ig}.

The present work is focused on the semileptonic decay of $D$ and $D_s$ mesons to the light scalar mesons namely, $f_0(980)$ and $a_0(980)$ in the framework of Covariant Confined Quark Model (CCQM) \cite{Efimov:1988,Efimov:1993,Ivanov:1999ic,Branz:2010}.
The CCQM is the effective field theory approach with the built-in infrared confinement for the hadronic interactions to their constituents.
Recently, we studied the semileptonic decays of $D$ and $D_s$ mesons to the pseudoscalar and vector mesons in this formalism in great detail \cite{Soni:2017eug,Soni:2018adu,Ivanov:2019nqd,Soni:2019qjs,Soni:2019huk}.
In these papers, we investigated the transition form factors, branching fractions and other physical observables such as forward backward asymmetry and lepton polarization.
The present study will help understand the essential dynamics of charmed semileptonic decays and the possible structure of the scalar mesons below 1 GeV namely, $f_0(980)$ and $a_0(980)$.

This paper is organised in the following way: after introducing the requirement for the study of scalar mesons in the semileptonic decays with the literature in Sec. \ref{sec:introduction}, we briefly introduce the essential components of Covariant Confined Quark Model employed here for computation of the hadronic form factors in Sec. \ref{sec:model}. Using the transition form factors, we compute the semileptonic branching fractions. In Sec. \ref{sec:results}, we present our numerical results of semileptonic branching fractions in comparison with other theoretical results and available experimental data. Finally, in Sec. \ref{sec:conclusion}, we summarize and conclude the present work.

\section{Form factors and semileptonic branching fractions}
\label{sec:model}
\begin{figure}[htbp]
\label{fig:semileptonic}
\includegraphics[width=0.5\textwidth]{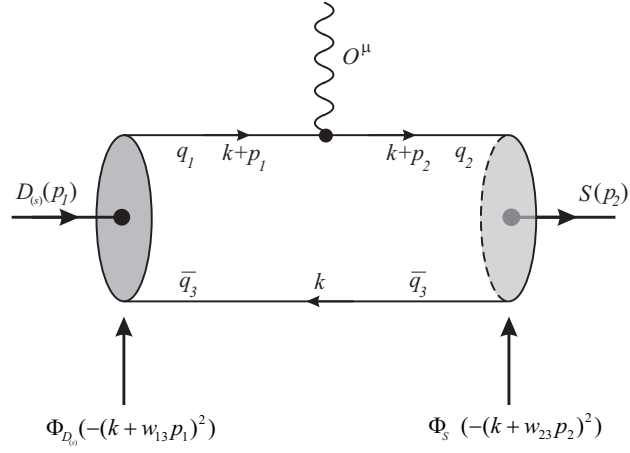}
\caption{Quark model diagrams for the $D$-meson leptonic decay}
\end{figure}

\begin{figure*}
\includegraphics[width=0.45\textwidth]{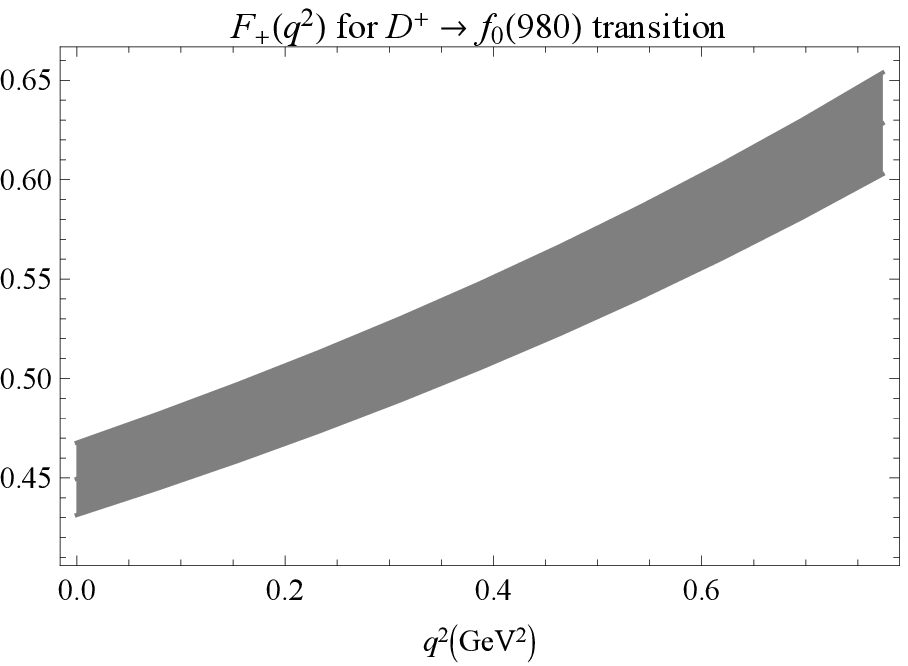}
\hfill\includegraphics[width=0.45\textwidth]{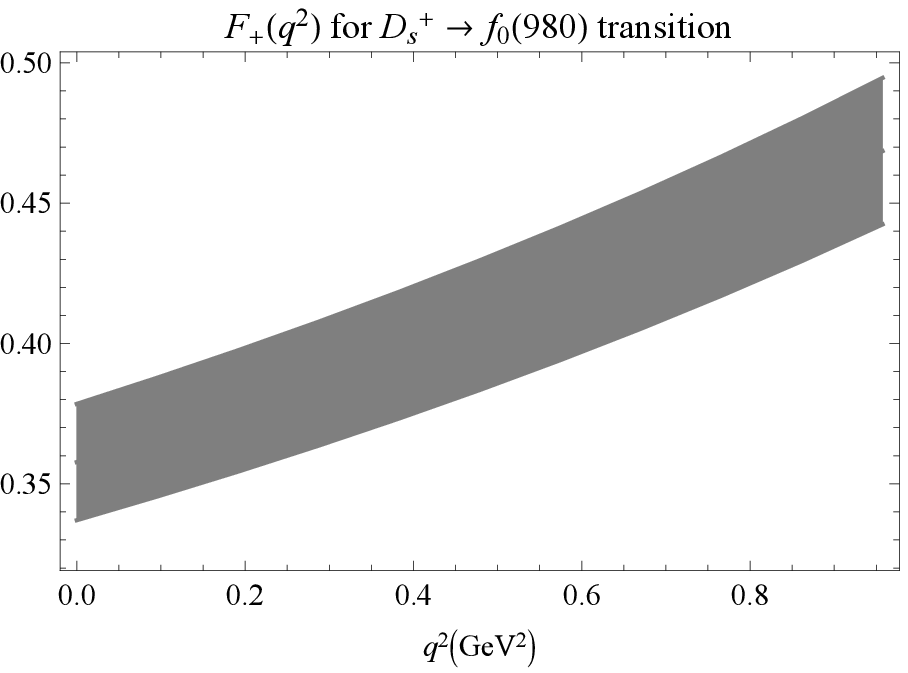}\\
\includegraphics[width=0.45\textwidth]{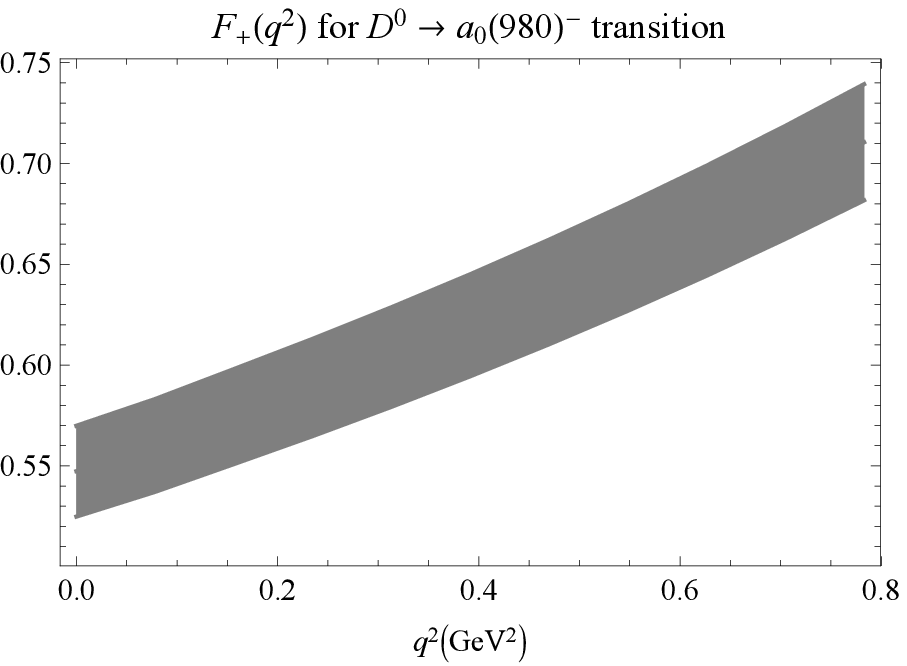}
\hfill\includegraphics[width=0.45\textwidth]{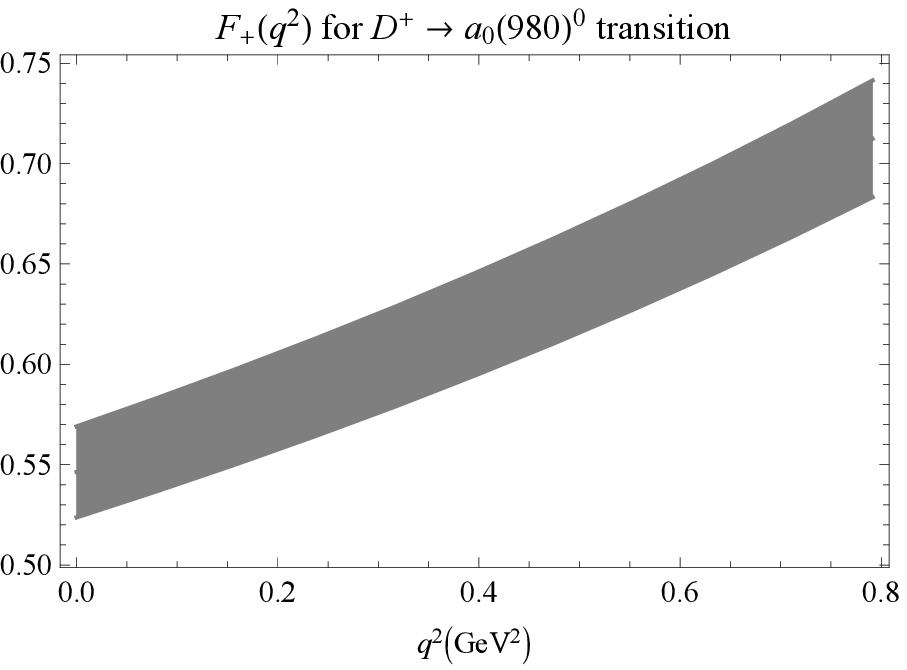}\\
\caption{$q^2$ dependence of the $D_{(s)} \to S$ form factors}
\label{fig:form_factors}
\end{figure*}

Within standard model, the semileptonic decays are very well separated by strong and weak interactions.  The charmed meson semileptonic decays to light scalar meson can be written as
\begin{eqnarray} \label{eq:int_lagrange}
\mathcal M(D_{(s)} \to S \ell^+ \nu_{\ell}) = \frac{G_F}{\sqrt{2}} V_{cq} \langle S | \bar{q} O^\mu c | D_{(s)} \rangle [\ell^+ O_\mu \nu_{\ell}], \nonumber
\end{eqnarray}
with $O^\mu = \gamma^\mu(1 - \gamma_5)$ and $q \in {d, s}$.
The matrix element in this process is very well parametrized in terms of transition form factors given by
\begin{eqnarray}
\mathcal{M}^{\mu}_S = P^{\mu} F_+(q^2) + q^{\mu} F_-(q^2)
\end{eqnarray}
Here $P = p_1 + p_2$ and $q = p_1 - p_2$ with $p_1$ and $p_2$ to be the momentum of $D_{(s)}$ meson of mass $m_1$ and momentum of Scalar ($S$) meson of mass $m_2$ respectively. The form factors $F_+$ and $F_-$ are computed in the entire accessible physical range of momentum transfer in the formalism of CCQM. The Lagrangian describing the coupling of the constituent quarks to the meson can be written as \cite{Efimov:1988,Efimov:1993,Ivanov:1999ic,Branz:2010}
\begin{dmath}
\mathcal{L}_{int}  =  g_M M(x) \int dx_1 dx_2 F_M(x;x_1,x_2) \bar{q}_2(x_2) \Gamma_M q_1(x_1) +  H.c
\end{dmath}
Here, $\Gamma_M$ is the Dirac matrix projecting onto spin of corresponding mesonic state. It should read i.e., $\Gamma_M = I, \gamma^5, \gamma_\mu$ for scalar, pseudoscalar and vector mesons respectively.
$g_M$ is the coupling strength of the meson with its constituent quarks. $F_M$, the translation invariant vertex function characterizing the effective physical size of the hadron, is given by
\begin{equation}\label{eq:vertex_function}
F_M(x,x_1,x_2) = \delta \left(x - \sum_{i=1}^2 w_i x_i \right)\Phi_M \left((x_1 - x_2)^2\right)
\end{equation}
with $\Phi_M$ as the correlation function of two constituent quarks with masses $m_{q_1}$ and $m_{q_2}$ and $w_{q_i} = m_{q_i}/ (m_{q_1} + m_{q_2})$ such that ${\mathit{w_1}} + {\mathit {w_2}} = 1$.
We choose Gaussian form for the vertex function as
\begin{equation} \label{eq:gaussian}
\tilde{\Phi}_M(-p^2) = \exp \ (p^2/\Lambda_M^2)
\end{equation}
where the model parameter $\Lambda_M$ characterizes the effective finite size of the mesons. Note that Eq. (\ref{eq:gaussian}) is the Fourier transform of the vertex function Eq. (\ref{eq:vertex_function}) for meson $M$.
The coupling strength $g_M$ can be determined using the renormalization of the one loop self energy Feynman diagram. This is also known as the compositeness condition which ensures the absence of any bare quark state in the final mesonic state \cite{Salam:1962,Weinberg:1962,Hayashi:1967bjx},
\begin{equation}
Z_M = 1 - \frac{3 g_M^2}{4 \pi^2} \tilde\Pi'_M(m^2_M) = 0.
\label{eq:Z=0}
\end{equation}
\begin{figure}[htbp]
\includegraphics{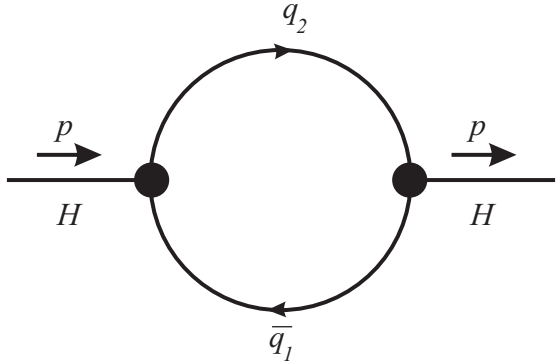}
\caption{Diagram describing meson mass operator}
\label{fig:mass_operator}
\end{figure}
The matrix element of self energy diagram and semileptonic decays are constructed from the $S$-matrix using the interaction Lagrangian Eq. (\ref{eq:int_lagrange}). The corresponding one loop Feynman diagram is drawn using the convolution of quark propagator and vertex functions (Fig. \ref{fig:mass_operator} and \ref{fig:semileptonic}).
The matrix element for self energy diagram for any meson can be written as
\begin{eqnarray}\label{eq:mass}
\tilde\Pi_M(p^2) &=& N_c g_{M}^{2}
\int  \frac{d^4k}{(2\pi)^4i} \tilde\Phi^2_M(-k^2) \cr && \times
\mathrm{tr}\Big(\Gamma_M S_1(k+w_1 p)\Gamma_M S_2(k-w_2 p)\Big).
\end{eqnarray}
In Eq. (\ref{eq:Z=0}), $\tilde\Pi^\prime_M$ is the derivative of meson mass operator Eq. (\ref{eq:mass}). Similarly, the matrix element for the semileptonic $D_{(s)}$ decays to scalar mesons can be written as
\begin{widetext}
\begin{eqnarray}
\label{eq:ff_DS}
\langle S (p_2) | \bar{q}O^\mu c  | D_{(s)}(p_1) \rangle  &=&
N_c g_{D_{(s)}} g_S \int \frac{d^4 k}{(2\pi)^4 i} \Phi_{D_{(s)}} (-(k + w_{13} p_1)^2) \Phi_S(-(k + w_{23} p_2)^2) \cr && \times \mathrm{tr}[S_2(k + p_2) O^\mu S_1(k + p_1) \gamma^5 S_3(k)]\cr
&=& F_+(q^2) P^{\mu} +  F_-(q^2) q^{\mu}
\end{eqnarray}
\end{widetext}
where $N_c = 3$ is the number of flavors.
The Fock-Schwinger representation of the quark propagator ($S_1$, $S_2$ and $S_3$) is used in computing the loop integral. This method involves the conversion of the loop momenta into the exponential of function.
The necessary loop integral can be evaluated analytically using the FORM code \cite{Vermaseren:2008kw}.
Finally, the universal infrared cutoff parameter $\lambda$ is used in computation that guarantees the quark confinement within the hadrons. We take $\lambda$ to be the same for all the physical processes. This computation technique is quite general and can be used for Feynman diagrams with any numbers of loops. All the numerical calculations including the multidimensional integrations are performed using a \textit{Mathematica} code developed by us.
For more detailed information regarding computation technique of the loop integral, we suggest the reader to refer to Refs. \cite{Branz:2010,Ivanov:2019nqd}.
The necessary model parameters for computation of semileptonic branching fractions are given in Tab. \ref{tab:parameters}.
These parameters have been determined for the basic electromagnetic properties like leptonic decay constants to match with the corresponding experimental data or lattice simulation results \cite{Ivanov:2011aa}.
For present computations, we employ model parameters that are obtained using updated least square fit procedure performed in the Refs. \cite{Ganbold:2014pua,Issadykov:2015iba,Ivanov:2019nqd}. The parametrization was achieved to keep the deviation in the computed decay constants defined by the function $\chi^2 =\sum_i \frac{(y_i^{\textrm{experiment}} - y_i^{\textrm{theory}})^2}{\sigma_i^2}$ to be minimum \cite{Ivanov:2016qtw,Ivanov:2017mrj,Tran:2018kuv}. Here, $\sigma_i$ are reported experimental standard deviations. After all the parameters were fitted to get the best possible decay constant values, the uncertainties in the model parameters were determined by individually changing them to get the exact experimental or lattice results. The difference between these two values of the parameters was considered as uncertainty in the respective parameter.  These uncertainties are considered absolute for given parameters and are then transported  to the form factors in the whole $q^2$ range. In Fig. \ref{fig:form_factors}, we present the spread of form factors $F_+$ in the whole $q^2$ range due to propagation of uncertainty in the parameters. It is observed that the uncertainties are of the order of $4~\% - 6~\%$ at the maximum recoil ($q^2 = 0$) and $8~\% -10~\%$ at the minimum recoil ($q^2 = q^2_{\textrm{max}}$).
Further, we compute their propagation in determination of uncertainty in branching fractions using the generic method given in the Appendix.
\begin{table}
\caption{Model parameters namely quark masses, size parameters and infrared cut off parameter (all in GeV).}\label{tab:parameters}
\begin{ruledtabular}
\begin{tabular}{ccccccccc}
$m_{u/d}$ & $m_s$ & $m_s$ & $\Lambda_D$ & $\Lambda_{D_s}$ & $\Lambda_{a_0}$ & $\Lambda_{f_0}^{q\bar{q}}$ & $\Lambda_{f_0}^{s\bar{s}}$ & $\lambda$\\
\hline
0.241 & 0.428 & 1.672 & 1.60 & 1.75 & 1.50 & 0.25 & 1.30 & 0.181
\end{tabular}
\end{ruledtabular}
\end{table}
\begin{table*}\caption{Double pole parameters for the computation of form factors in Eq. \ref{eq:double_pole}}
\label{tab:double_pole_data}
\begin{ruledtabular}
\begin{tabular}{lccclccc}
$F$ & $F(0)$ & $a$ & $b$ & $F$ & $F(0)$ & $a$ & $b$\\
\hline
$F_+^{D^+ \to f_0(980)}$ 		& $0.45 \pm 0.02$ & 1.36 & 0.32 & $F_-^{D^+ \to f_0(980)}$ 	& $0.40 \pm 0.02$ 	& 0.71 & 0.24\\
$F_+^{D_s^+ \to f_0(980)}$ 	& $0.36 \pm 0.02$ & 0.99 & 0.13 & $F_-^{D_s^+ \to f_0(980)}$ 	& $-0.39 \pm 0.02$	& 1.13 & 0.18\\
$F_+^{D^0 \to a_0(980)^-}$ 	& $0.55 \pm 0.02$ & 1.05 & 0.15 & $F_-^{D \to a_0(980)}$ 		& $0.03 \pm 0.01$ & $-$0.04 & 32.81\\
 $F_+^{D^+ \to a_0(980)^0}$ & $0.55 \pm 0.02$ & 1.06 & 0.16 & $F_-^{D \to a_0(980)}$ 		& $0.03 \pm 0.01$	& 1.43 & 72.93
\end{tabular}
\end{ruledtabular}
\end{table*}

\begin{table}
\caption{Comparison of the form factor at the maximum recoil}
\label{tab:form_factor_comparison}
\begin{ruledtabular}
\begin{tabular}{cccl}
Channel & Present & Other & Ref.\\
\hline
$D^+ \to f_0(980)$ 	& 0.45 $\pm$ 0.02	& 0.321 					& LCSR \cite{Shi:2017pgh} \\
								&							& 0.216						& LFQM \cite{Ke:2009ed}\\
								& 							& 0.21						& CLFD \cite{ElBennich:2008xy}\\
								& 							& 0.22						& DR \cite{ElBennich:2008xy}\\
$D_s^+ \to f_0(980)$ & 0.39 $\pm$ 0.02 & 0.30 $\pm$ 0.03 & LCSR \cite{Colangelo:2010bg}\\
								&							& 0.434						& LFQM \cite{Ke:2009ed}\\
								& 							& 0.45						& CLFD \cite{ElBennich:2008xy}\\
								& 							& 0.46						& DR \cite{ElBennich:2008xy}\\
$D^0 \to a_0(980)^-$	& 0.55  $\pm$ 0.02& $1.75^{+0.26}_{-0.27}$ & LCSR \cite{Cheng:2017fkw}\\
$D^+ \to a_0(980)^0$	& 0.55  $\pm$ 0.02& $1.76 \pm 0.26$	& LCSR \cite{Cheng:2017fkw}
\end{tabular}
\end{ruledtabular}
\end{table}

The form factors given in Eq. (\ref{eq:ff_DS}) are also very well represented in the double pole approximation as
\begin{eqnarray}\label{eq:double_pole}
F(q^2) = \frac{F(0)}{1 - a s + b s^2}, \ \ \ \ s = \frac{q^2}{m_{D_{(s)}}^2}
\end{eqnarray}
The parameters in the double pole approximation for the different decay channels are given
in the Tab. \ref{tab:double_pole_data}. It is worth mentioning that the parametrization in the double pole approximation is quite precise and the deviation of the form factors from the actual data is less than 1 \% in the entire range of momentum transfer.

Using the necessary model parameters (Tab. \ref{tab:parameters}) and computed form factors (Tab. \ref{tab:double_pole_data}), we determine the semileptonic branching fractions in terms of helicity structure functions using the relation \cite{Gutsche:2015mxa,Ivanov:2015tru}
\begin{dmath}\label{eq:decay_width}
\frac{d\Gamma(D_{(s)} \to S \ell^+ \nu_\ell)}{dq^2} = \frac{G_F^2 |V_{cq}|^2 |{\bf p_2}| q^2 v^2}{12 (2\pi)^3m_1^2} ((1+\delta_\ell) |H_0|^2 + 3 \delta_\ell |H_t|^2)
\end{dmath}
where $\delta_\ell = m_\ell^2/2q^2$ is the helicity flip factor, $|{\bf p_2}| = \lambda^{1/2} (m_{D_{(s)}}^2, m_S^2, q^2)/2 m_{D_{(s)}}$ is the momentum of the daughter (Scalar) meson in the rest frame of the parent ($D_{(s)}$) meson and $v = 1-m_\ell^2/q^2$ is the velocity-type parameter. In the above Eq. (\ref{eq:decay_width}), the bilinear combinations of the helicity structure function are defined in terms of form factors as:
\begin{eqnarray}\label{eq:helicity}
H_t &=& \frac{1}{\sqrt{q^2}} (Pq F_+ + q^2 F_-),\nonumber \\
H_0 &=& \frac{2 m_1 |\bf{p_2}|}{\sqrt{q^2}} F_+
\end{eqnarray}
This helicity technique is formulated in Refs. \cite{Korner:1987kd,Korner:1989qb,Korner:1989ve} and is also discussed recently in Refs. \cite{Gutsche:2015mxa,Ivanov:2015tru}. The computation technique in CCQM is very general and can accommodate  hadronic state with any number of constituent quarks.

\section{Numerical results and Discussion}
\label{sec:results}
After fixing the quark masses and meson size parameters, we compute the transition form factors for the semileptonic decays of $D$ and $D_s$ mesons to the scalar mesons below 1 GeV in the entire physical range of momentum transfer.
Looking at the literature, theoretically the internal structure of these scalar mesons are still not very much clear.
In this study, we consider the internal structure of scalar meson $a_0 (980)$ to be the conventional quark antiquark state while $f_0(980)$ meson as the admixture of $q\bar{q}$ and $s\bar{s}$ state characterized by a mixing angle $\theta$. In terms of wave-function, the internal structures of these scalar mesons are defined as
\begin{eqnarray*}
&&|f_0(980)\rangle = \cos \theta |s\bar{s}\rangle + \sin\theta |q\bar{q}\rangle  \\
&&|a_0(980)^-\rangle =  |d\bar{u}\rangle, \\
&&|a_0(980)^0\rangle  =  \frac{1}{\sqrt{2}}  |u\bar{u} - d\bar{d}\rangle
\end{eqnarray*}
with $q\bar{q} = 1/\sqrt{2}(u\bar{u} + d\bar{d})$ and the mixing angle $\theta$ to be in the range $25^\circ < \theta < 40^\circ$ and $140^\circ < \theta < 165^\circ$ \cite{Cheng:2002ai,Wang:2006ria}.
Similar approach has been used earlier in the formalism of light front quark model and the authors obtained the mixing angle to be $(56 \pm 7)^\circ$ or $(124 \pm 7)^\circ$ \cite{Ke:2009ed}.
Another approach was used in covariant quark model formalism by B. E. Bennich \textit{et al} where they considered the  phenomenological analysis of two body decay of $D_{(s)}$ mesons and the form factors were computed using the dispersion relation (DR) and covariant light front dynamics (CLFD) \cite{ElBennich:2008xy}.
Furthermore, various other approaches have been reported considering the quark contribution. For example, E. Oset \textit{et al}. have studied the structure of $f_0(980)$ in various channels using chiral unitary approach and inferred that the $f_0(980)$ has major contribution from the strange quarks \cite{Sekihara:2015iha,Oller:1998hw,Oller:1997ng,Oller:1997ti,Oller:1998zr}. L. Bediaga \textit{et al}., have studied the structure of $f_0(980)$ in semileptonic decay $D_{(s)} \to f_0(980) \ell^+ \nu_\ell$ using QCD sum rules approach and inferred that the contribution of light quark component is not negligible \cite{Bediaga:2003zh}.
Motivated by these findings clearly suggesting considerable strange quark contribution, we choose the mixing angle to be in the range $25^\circ < \theta < 40^\circ$ \cite{Cheng:2002ai,Wang:2006ria} for present computations and the computed branching fraction range turns out to be
\begin{eqnarray}
\mathcal{B}(D^+ \to f_0(980) e^+ \nu_e) & = & (5.95 - 10.06) \times 10^{-5} \nn
\mathcal{B}(D^+ \to f_0(980) \mu^+ \nu_\mu) & = & (5.92 - 10.01) \times 10^{-5} \nn
\mathcal{B}(D_s^+ \to f_0(980) e^+ \nu_e) & = & (0.14 - 0.27) \times 10^{-2} \nn
\mathcal{B}(D_s^+ \to f_0(980) \mu^+ \nu_\mu) & = & (0.14 - 0.26) \times 10^{-2} \nonumber
\end{eqnarray}
In Tab. \ref{tab:branching}, we quoted the central values of branching fractions and the corresponding value of the mixing angle is $31^\circ$. Also, in Tab. \ref{tab:double_pole_data} and \ref{tab:form_factor_comparison}, we quoted the values of form factors and double pole parameters for mixing angle of $31^\circ$.
In computing the transition form factors, we have considered contribution from $q\bar{q}$ for the channel $D \to f_0(980)$ meson and that from $s\bar{s}$ for $D_s \to f_0(980)$.
The form factors appearing in Eq. (\ref{eq:ff_DS}) are computed in the entire accessible range of momentum transfer and associated double pole parameters Eq. (\ref{eq:double_pole}) are tabulated in Tab. \ref{tab:double_pole_data}.
We also compare our results of the form factors at the maximum recoil ($q^2 = 0$) in Tab. \ref{tab:form_factor_comparison} along with the other theoretical models using light cone sum results data, covariant light front dynamics and dispersion relations.
Our results of $f_+(0)$ for $D \to f_0(980)$ are lower than those obtained using the LCSR \cite{Shi:2017pgh} from the theoretical analysis of $D \to \pi\pi \ell \bar{\nu}$ decays. However, they match well with the quark-antiquark picture of scalar meson in LFQM \cite{Ke:2009ed} and the mixing of $s\bar{s}$ with light quark state of scalar mesons in CLFD/DR \cite{ElBennich:2008xy} approach.
For $D_s \to f_0(980)$ channel, our result of the $f_+(0)$ matches well with the LCSR \cite{Colangelo:2010bg} but it is lower than the LFQM and CLFD/DR approach.
We also provide the form factors for the channel $D \to a_0(980)$ in comparison with the LCSR results \cite{Cheng:2017fkw} where the structure of $a_0(980)$ is considered to be the conventional $q\bar{q}$ state.

The computed form factors are then utilized for calculation of semileptonic branching fractions using Eq. (\ref{eq:decay_width}).
Our results of semileptonic branching fractions for both electron channel and muon channel are presented in Tab. \ref{tab:branching} in comparison with other theoretical and available experimental data. No experimental results are available for the absolute branching fractions of $D_{(s)}^+ \to f_0(980) e^+ \nu_e$. However, recently BESIII have set the upper limit on the $\mathcal{B}(D^+ \to f_0(980) e^+ \nu_e, f_0(980) \to \pi^+ \pi^-) < 2.8 \times 10^{-5}$ with the confidence level of 90~\% \cite{Ablikim:2018qzz}. Also Hietala \textit{et al}., predicted $\mathcal{B}(D_s^+ \to f_0(980) e^+ \nu_e, f_0(980) \to \pi^+ \pi^-) = (0.13 \pm 0.03 \pm 0.01)~\%$ using the CLEO-c data \cite{Hietala:2015jqa}.
For $D \to a_0(980)$ channel also, the absolute value of branching fraction is not available in the BESIII paper \cite{Ablikim:2018ffp}. They have predicted the ratio of the partial width $\frac{\Gamma(D^0 \to a_0(980)^-e^+\nu_e)}{\Gamma(D^+ \to a_0(980)^0 e^+ \nu_e)} = 2.03 \pm 0.95 \pm 0.06$ and we obtained the ratios to be $1.95 \pm 0.38$ which is within the uncertainty limits predicted by them \cite{Ablikim:2018ffp}.

\begin{table*}
\caption{Branching fractions of $D \to S$ semilleptonic decay}
\label{tab:branching}
\begin{ruledtabular}
\begin{tabular}{lcccc}
Channel &Unit & Present &Theory & Reference \\
\hline
$D^+ \to f_0(980) e^+ \nu_e$ 			& $10^{-5}$	& 7.78 $\pm$ 0.68 & 5.7 $\pm$ 1.3 & LFQM \cite{Ke:2009ed}\\
$D^+ \to f_0(980) \mu^+ \nu_\mu$ 		& $10^{-5}$ 	& 7.87 $\pm$ 0.67 \\
$D_s^+ \to f_0(980)e^+\nu_e$ 			& $10^{-2}$ 	& 0.21 $\pm$ 0.02 & $0.2^{+0.05}_{-0.04}$ & LCSR \cite{Colangelo:2010bg} \\
$D_s^+ \to f_0(980)\mu^+\nu_\mu$	& $10^{-2}$ 	& 0.21 $\pm$ 0.02\\
\hline
$D^0 \to a_0(980)^-e^+ \nu_e$			& $10^{-4}$ 	& 1.68 $\pm$ 0.15  & $4.08^{+1.37}_{-1.22}$ & LCSR \cite{Cheng:2017fkw} \\
$D^0 \to a_0(980)^-\mu^+ \nu_\mu$	& $10^{-4}$ 	& 1.63 $\pm$ 0.14 \\
$D^+ \to a_0(980)^0e^+ \nu_e$			& $10^{-4}$	& 2.18 $\pm$ 0.38  & $5.40^{+1.78}_{-1.59}$ & LCSR \cite{Cheng:2017fkw} \\
$D^+ \to a_0(980)^0\mu^+ \nu_\mu$	& $10^{-4}$	& 2.12 $\pm$ 0.37 \\
$\frac{\Gamma(D^0 \to a_0(980)^-e^+ \nu_e)}{\Gamma(D^+ \to a_0(980)^0e^+ \nu_e)}$ & -- & 1.95 $\pm$ 0.38 &&
\end{tabular}
\end{ruledtabular}
\end{table*}

\section{Conclusion}
In this paper, we have considered the $f_0(980)$ meson to be the admixture of $s\bar{s}$ and light quark component with the mixing angle of $25^\circ < \theta < 40^\circ$ and $a_0(980)$ meson to be the conventional quark-antiquark pair. We have employed the Covariant Confined Quark Model to study the semileptonic branching fraction of charmed mesons decaying to the light scalar mesons. Our results are found to be consistent with theoretical results as well as available experimental data.
The present study indicates that the internal structure of $f_0(980)$ has higher contribution of $s\bar{s}$ state suggesting the validity of chiral unitary approach, light cone sum rules analysis and light front quark model approaches.
We have also provided theoretical prediction for the semileptonic branching fractions of charmed meson to scalar meson in the muon channel for the first time.

The present study can help in understanding the internal structure of the scalar mesons below 1 GeV. As no absolute value of the branching fractions is available in the literature, we expect more accurate data coming from the worldwide upgraded experimental facilities to check the validity of computed results in this study.
\label{sec:conclusion}

\section*{ACKNOWLEDGMENTS}
We thank Prof. Mikhail A. Ivanov for the continuous support throughout this work and for providing critical remarks for improvement of the paper.
J.N.P. acknowledges financial support from University Grants Commission of India under Major Research Project F.No. 42-775/2013(SR).
N.R.S. and A.N.G.  thank Bogoliubov Laboratory of Theoretical Physics, Joint Institute for Nuclear Research for warm hospitality during Helmholtz-DIAS International Summer School ``Quantum Field Theory at the Limits: from Strong Field to Heavy Quarks” where this work was initiated.
%\clearpage

\appendix
\section{Propagation of Uncertainty}
\label{Appen:error}
The error propagation in the branching fraction can be computed using the most common technique. In general, the differential branching fractions Eq. (\ref{eq:decay_width}) can also be rewritten in terms of form factors as
\begin{eqnarray}
\frac{d\mathcal{B}}{dq^2} = N (a F_+^2(q^2) + b F_-^2(q^2) + c F_+(q^2) F_- (q^2))
\end{eqnarray}
Here $N$ includes the terms involving the various constants such as Fermi coupling constant, CKM matrix elements, meson masses etc and $a, b$ and $c$ are the coefficients of form factors in Eq. (\ref{eq:helicity}).
The uncertainty in the measurement of branching fractions because of the uncertainty in form factors can be written as
\begin{equation}\label{eq:error_br}
d (\Delta \mathcal{B}) = dq^2 N \sqrt{\left(\frac{\partial (d\mathcal{B})}{\partial F_+} \Delta F_+\right)^2 + \left(\frac{\partial (d\mathcal{B})}{\partial F_-} \Delta F_-\right)^2}
\end{equation}
where $(\Delta F_i)^2 = (F_i \cdot \varepsilon_i)^2$ with $\varepsilon$ is the relative error of all the form factors. The uncertainties in the form factors are also extracted using the same method. Finally we determine the uncertainty in the branching fractions by integrating the above equation.

\bibliography{apssamp2}% Produces the bibliography via BibTeX.

\end{document}